# Framework for Visualizing Model-Driven Software Evolution and its Application

Akepogu ANAND RAO and Karanam MADHAVI

Computer Science & Engineering Department
Jawaharlal Nehru Technological University,
Anantapur, Andhra Pradesh, India.

**Abstract**
Software Visualization encompasses the development and evaluation of methods for graphically representing different aspects of methods of software, including its structure, execution and evolution. Creating visualizations helps the user to better understand complex phenomena. It is also found by the software engineering community that visualization is essential and important. In order to visualize the evolution of the models in Model-Driven Software Evolution, authors have proposed a framework which consists of 7 key areas (views) and 22 key features for the assessment of Model Driven Software Evolution process and addresses a number of stakeholder concerns. The framework is derived by the application of the Goal Question Metric Paradigm. This paper aims to describe an application of the framework by considering different visualization tools/CASE tools which are used to visualize the models in different views and to capture the information of models during their evolution. Comparison of such tools is also possible by using the framework.
***Keywords:*** *Model-Driven Software Evolution, Software Visualization, Visualization tools.*

## 1. Introduction

Visualization is used to enhance information understanding by reducing cognitive overload. Using visualization methodologies and tools, people are often able to understand the information presented in a shorter period of time or to a greater depth. The term "visualization" can refer to the activity that people undertake when building an internal picture about real world or abstract entities. Visualizing can also refer to the process of determining the mappings between abstract or real-world objects and their graphical representation. This work uses the term "visualization" in the later sense: the process of mapping
the evolution of models to the stakeholder concerns.

The introduction of Model Driven Engineering (MDE) needs a new style of evolution i.e. Model-driven Software Evolution. The first fundamental premise [1] for Model-Driven Software Evolution (MoDSE) is that evolution should be a continuous process. The second premise is that reengineering of legacy systems to the model-driven of the paradigm should be done incrementally. MDE introduces a multitude of languages that are themselves artifacts of the development process. Due to these multitude languages in MoDSE, there is a need to have the model interaction, integration, mapping and transformation. Further there should be possible views to capture this information about models during the evolution. For this purpose multiple views for MoDSE have been proposed in [9]. Stakeholder's involvement in MoDSE typically has interests in, or concerns relevant to that system. The ability of models to evolve gracefully is becoming a concern for many stakeholders. Due to different and interrelated models used to design an entire system in MoDSE, the concerns of stakeholders may differ from one role to another role that a stakeholder play during the life time of a software project. So, visualization provides better solution to understand the complex information during evolution of the models. This can be done by using the existing visualization and/or CASE tools. Software Visualization tools use graphical techniques to make software artifacts visible.

Evaluating a particular visualization tool for MoDSE is essential. Common practice is that some set of guidelines are followed and a qualitative summary is produced. However, these guidelines do not usually allow a comparison of competing techniques or tools. A comparison is important because it identifies possible flaws in the research area or software development. Thus, a framework for describing attributes of tools is needed. Once the tools have been assessed in this common framework, a comparison is possible. However, a framework can be used for comparison, discussion, and formative evaluation of the tools. Such framework was proposed in [8]. So, the major contribution of this paper is to show how the framework can be applied to compare the Visualization Tools which is presented in section 4. A Framework for visualizing Model Driven Software Evolution falls into seven key areas (views): Context View, Inter-model View, City View, Metric View, Transformation View, Evolution View and Evaluation view [9] and 22 Key features are identified for all key areas. The framework is used to evaluate visualization tools and it is also used to assess tool appropriateness from a variety of stakeholder perspectives.

This paper is structured as follows: Section 2 discusses the related work. Section 3 summarizes the framework. Section 4 discusses an application of the framework by considering different Visualization tools/CASE tools. Section 5 outlines the conclusions and giving an outlook on future work.

## 2. Related Work

This section reviews the literature related to the fields of Software Visualization, Software Evolution Visualization and Model Driven approaches.

Source Viewer 3D (sv3D) [6] is a Software Visualization framework that builds on the SeeSoft metaphor. sv3D can show large amounts of source code in one view. Object based manipulation methods and simultaneous alternative mappings are available to the user. The types of user tasks and interactions that are supported by sv3D, is not directly related to solving/visualizing specific software engineering tasks and it is a prerequisite for a software visualization tool.

Architecture to Support Model Driven Software Visualization [7], borrows the field of Model Driven Engineering (MDE) to assist with the creation of highly customizable interfaces for Software Visualization. In order to validate the architecture, MDV framework for Eclipse was developed. Model Driven Visualization (MDV) is intended to address the customization of information visualization tools, especially in the program comprehension domain. The MDV architecture describes how to leverage the work done in the Model Driven Engineering community and apply it to the problem of designing visualizations tools.

The Graphical Modeling Framework(GMF)[12] project for
eclipse has facilities to allow modelers to define graphical editors for their data. These graphical editors can be used as viewers, however, the views they support are limited to simple graphs with containers. The GMF project currently lacks the ability to specify "Query Result" visualizations.

An Open Framework for [10] visual mining of CVS based software repositories has three major aspects are data extraction, analysis and visualization. An approach was proposed for CVS data extraction and analysis. CVS data acquisition mediator used to extract the data from CVS repositories. Analysis techniques are used to analyze the raw data retrieved from the CVS repositories from CVS Querying. It also provides the comparison of the open source projects. CVSgraph is a software tool used to visualizing project at file level. This open framework does not provide the visualization of models, it provides for program at file level only.

CVSscan[11] is a tool in which a new approach for visualization of software evolution was developed. The main audience targeted here is the software maintenance community. The main goal is to provide support for program and process understanding. This approach uses multiple correlated views on the evolution of a software project. The overall evolution of code structure, semantics, and attributes are integrated into an orchestrated environment to offer detail-on-demand. And also provides the code text display that gives a detailed view on both the composition of a fragment of code and its evolution in time. It is focused on the evolution of individual files.

2.1 Motivation for Framework and its Application

There are number of frameworks exists in the literature for comparison and assessment of the various CASE tools. Comparison of these tools is essential to understand their differences, to ease their replication studies, and to discover what tools are lacking. Such a comparison is difficult because there is no well-defined comprehensive and common comparative study for different category of the tools. For design recovery tools a comparative framework [14] was derived for comparison. This framework comprises eight concerns, which were further divided into fifty three criteria and which were applied on ten design recovery tools successfully. Another framework [7] also exists in the literature for comparison and assessment of the software architecture visualization tools. Software architecture is the gross structure of a system; as
such, it presents a different set of problems for visualization than those of visualizing the software at a lower level of abstraction. Six visualization tools were evaluated in this framework. This framework consists of seven Key areas and 31 Key features, for the assessment of software architecture visualization tools. Both the frameworks applied against the stakeholders concerns. From this analysis it is easy to know that how a selected tool satisfies the stakeholder concerns. Thus the motivation for this work lies in above mentioned two frameworks. Combining the visualization approach with MoDSE is essential to understand the evolution of models in a better way. Large numbers of visualization tools are available in the literature. Among them many tools support the evolution at source code level, data level. This work aims to find out the visualization tools which support the visualization at model level. As such there is no framework exists in the literature to evaluate tools which are useful for the MoDSE Visualization and also to understand the evolution of the models with respect to stakeholder perspectives. Hence this paper aims to evaluate the already proposed framework for visualization of MoDSE.

## 3. Framework Summary

This section provides the summary of the already proposed framework for Model-Driven Software Evolution visualization in [8].

The framework has seven key areas (views) for visualizing MoDSE: Context View, Inter-Model View, City View, Metric View, Transformation View, Evolution View and Evaluation view. These seven views are derived based on the viewpoints and were discussed in detail [8]. The dimensions proposed in the framework are not proposed as formal representation of the characteristics of MoDSE, but are necessary for discussion about, and evaluation of, such dimensions with respect to stakeholders and tools which they use. The Goal/Question/Metric (GQM) paradigm [9] is used to identify the questions and then to enable the formation of framework features.

The primary goal of the framework is to assess and understand the evolution of the models in model driven software evolution. The framework is derived from an extensive analysis of the literature in the area of software visualization with special emphasis on model driven software evolution. Each of the seven views is a conceptual goal which the framework must satisfy. It is this that makes the application of the GQM Paradigm [13] straightforward.

Framework summary with its goals, questions are given in Table.1 First column represents the key features (questions) which are abbreviated with view names. Second column represents the key areas (views). The responses for these questions will be the values used in the Table 2.

Table 1: Framework Summary

| Key Features | Key Areas |
|---|---|
| | **Key Area 1 :** Context View (CV) |
| CV 1 | Does the visualization provide context of a model? |
| CV 2 | Does the visualization provide the scope of a model or model element? |
| CV 3 | Does the visualization express the model completely including all its surrounding elements? |
| | **Key Area 2:** Inter-Model View (IMV) |
| IMV 1 | Does the visualization provide the dependencies between the models and model elements? |
| IMV 2 | Does the visualization provide the indirect dependencies between the models and model elements? |
| IMV 3 | Does the visualization provide the integration of the two or more models? |
| | **Key Area 3** : City View (CiV) |
| CiV 1 | Does the visualization provide the extendibility of the models in a software system? |
| CiV 2 | Does the visualization provide the traceability of a model or model element? |
| | **Key Area 4** : Metric View (MeV) |
| MeV 1 | Does the visualization provide the metrics to estimate the impact analysis of the models during evolution? |
| MeV 2 | Does the visualization provide the visualization techniques to know the evolution of the models? |
| MeV 3 | Does the visualization provide the metric values to know the evolution of the models? |
| MeV 4 | Does the metrics provide the knowledge about the quality and complexity of the models during evolution? |
| | **Key Area 5** : Transformation View (TV) |
| TV 1 | Does the visualization provide any kind of transformation? |
| TV 2 | Does the visualization provide the knowledge about the transformation of the models? |
| TV 3 | Does the visualization provide the mapping of models? |
| TV4 | Does it provide transformation rules? |
| TV5 | Does it provide transformation language? |
| | **Key Area 6** : Evolution View (EV) |
| EV 1 | Does the visualization provide trends and causes for evolution of models? |
| EV 2 | Does the visualization provide the dimension of evolution? |
| | **Key Area 7** : Evaluation View (EaV) |
| EaV 1 | Does the visualization provide the evolution trends and techniques? |
| EaV 2 | Does the visualization causes for the evolution of models? |
| EaV 3 | Does the visualization facilitate the stakeholders' feedback? |

## 4. Application of the Framework

This section describes an application of the framework. For this purpose tools which are mainly research oriented and non commercial tools are considered. These tools are also having the features which are necessary for visualization of models. The expensive commercial tools such IBM rational Rose Suite, Enterprise architect etc. are not considered here. The following sub sections briefly describe the features of the tools. And the comparison of those tools and responses are shown in the Table 3.

Table 2: Possible Responses (Metrics)

| Response | *Meaning* |
|---|---|
| Y | Full support |
| Y? | Mainly supported |
| N? | Mainly not supported |
| N | No support |
| NA | Not applicable (not in the scope) |
| ? | Unable to determine |

4.1 Argo UML Tool

ArgoUML (ARUML) is a free UML diagramming tool [5], [17] released under the open source BSD License. It is a java based UML tool that helps users to design using UML. It is able to create and save most of the nine standard UML diagrams. ArgoUML not only a free UML modeling tool, it is also an open source project that any one can contribute to extend or to customize the features of a tool. It is a powerful yet easy-to-use interactive, graphical software design environment that supports the design, development and documentation of Object-Oriented software applications. The users of Argo UML are software designers, architects, software developers, business analysts, system analysts and other professionals involved in the analysis, design and development of software applications. First version released in April 1998 and the recent version is 0.26.2 in November 2008.All nine UML 1.4 diagrams supported and it also supports many features but the major weakness is no support for UML 2. The four key features that make ArgoUML different from other tools are: it makes use of ideas from cognitive psychology, it is based on open standards and it is 100% pure java is used.

Explorer View in ArgoUML has 9 perspectives which satisfy the features of the framework such as CV1, CV2, CV3, IMV1, and IMV2. This is indicated with the response 'Y' in the Table 3. Integration of the models (IMV3) is not supported so, the response is 'N'. Features such as CiV1, CiV2 are not mainly supported because as such there is no geographical view of a complete project but it provides all the models in a project in a hierarchal tree view. Hence the response is 'N?' in the Table 3. All the features in a (Mev1, MeV2, MeV3, and MeV4) Metric View are not applicable here because it is not intended to calculate the metric values of the models. This is shown as 'NA' response. The response for the Transformation features such as TV1, TV2, and TV3 is 'N?' because transformation from model to code is partially available not the other kinds of transformation such as model to model or code to model. Transformation rules and language (TV4 and TV5) is not applicable, so the response is 'NA'. Compare to other two tools, ArgoUML is particularly inspired by the three theories within Cognitive Psychology. So, the designers of a complex system do not conceive a design fully formed. Instead, they must construct a partial design, evaluate, reflect on, and revise it, until they are ready to extend it further. So, the responses for the features are shown in the Table 3 as EV1 – Y? , EV2 – N, EaV1 – N, Eav2 – Y? , EaV3 – Y.

4.2 MetricView Evolution Tool

MetricView Evolution (MVE) tool [2], [3], [15] is a research activity within Empirical Analysis of Architecture and Design Quality Project (EmpAnAda). This Project is an activity of the System Architecture and Networking group at the Eindhoven University of Technology, Netherlands. MetricView Evolution tool provides features such as metrics calculations within the tool, several views to explore and navigate UML models, visualization of evolution data. This is an extension of MetricView tool which includes more features. MetricView Evolution also supports analysis of model quality and model evolution. Due to some limitations in this research activity and since the entire UML specification is quite complex so, not all the information available in each diagram. Only the necessary elements are extracted and displayed in this tool. Even with limitations the reasons to select this tool is research activity, easily downloadable and features are closer to the framework.

MetricView Evolution tool has full support (Y) for the key features such as CV1, CV2, CV3, IMV1, IMV2, CiV1, CiV2, MeV1, MeV2, MeV3, EV1, EaV1, and EaV2. Feature IMV3 (i.e. integration of models) is not supported but there is a scope for integrating the models. Features such as TV1, TV2, and TV3 are not applicable because these features are not in the scope of the tool. And the purpose of the MetricView Evolution tool is for quality and evolution of UML models not for transformation of models like model to model, code to model and model to code. EV2 and EaV2 features are not mainly supported (N?) because the purpose of the evolution view in the tool is to enable the user to spot the trends in the values of quality attributes and/or metrics at multiple abstraction levels not for multiple dimensions of evolution. The responses for the stakeholders concerns(key features or questions) are shown in the Table 3 in terms of Y, N, Y?, N? and NA.

4.3 Visual Paradigm for UML

Visual Paradigm [16] for UML 6.4 (VP-UML) is a powerful visual UML CASE tool. It is designed for a wide range of users, including software engineers, system analysts, business analysts, and system architects like who are interested in building software systems reliably through the use of Object-Oriented approach. VP-UML can run in different operating systems. It supports more than 20 diagram types including UML 2.1, BPMN, SysML, ERD, DFD and more. Different editions are also available such as Enterprise, Professional, Standard, Modeler, and Personal are commercial editions. Community and Viewer are non-commercial editions. It supports a rich array of tools. One special feature is Resource-Centric interface, which lets the user access modeling tools easily without referring back and forth from the workspace to various toolbars. Users can draw diagrams or models as with a pen and paper, executing complicated modifications with just a click and drag, creating completely visual environment

It is observed that the names of the features in VP-UML differ from the features of the framework. But the purpose and intention of the features are same. So, they have full support for those features that labeled as 'Y' in the Table 3. Transformation of the models such as model to model, model to code and code to model available in the tool but transformation rules and languages are not available. Hence, features as TV4, TV5 are not applicable (NA). MeV1,MeV2, MeV3, MeV4 features for metrics of the models and which are not in the VP-UML tool that is shown in the Table 3 as ' NA'. Features such as EV2, EaV3 are not mainly supported in the tool i.e. shown in the Table 3 as 'N?' Visualization of the models by using different diagrams is possible but the techniques are not available. So, the response is 'N?' for Mev2. Stakeholder's feedback (EaV3-N?) is not mainly provided, but the user can store their opinions/ideas about the evolution of the models.

## 5. Conclusions and Future work

An application of the framework for visualizing Model-Driven Software Evolution has been presented. The research oriented, non commercial tools such as ArgoUML, MetricView Evolution, and Visual Paradigm for UML are considered for the framework's application. These three tools have compared successfully under this common framework. From this comparison it is observed that a single tool does not consist of all the features of the framework and each tool has its own intensions and purposes. But, by using these three tools all the features are satisfied except four features. Among these two features such as 'multiple dimensions of evolution', 'stake holder's feedback' are partially supported by the two tools. But, ArgoUML has provided the feature such as 'Cognitive Psychology' which provides freedom for a stakeholder (designer) to make design decisions, to resolve design problems and many design issues and rules is also available. The remaining two features like 'transformation rules' and 'transformation languages' are not applicable and not supported by these three tools. By comparing the tools under this common framework a stakeholder can easily understand and asses the tools and can find out the flaws in a particular tool.

From the comparison of various features of the three tools it is observed that still there is a need to consider few more possible visualization/CASE tools which are exists in the literature. It is possible to check the unsatisfied features of the three tools can be satisfied by the other tools and also possible to know the role of the visualization tools in MoDSE. From the comparison of number possible tools framework can be strengthen further. Another application of the framework is to evaluate stakeholder concerns considered in the framework against the concerns of the software practitioners (stakeholders) from diverse organizations. These are the subjects of the future work.

Table 3: Framework Application

| Key Features | Key Areas | ARUML | MVE | VPUML |
|---|---|---|---|---|
| | *Context View (CV)* | | | |
| CV1 | Context of a model | Y | Y | Y |
| CV2 | Scope of a model or model element | Y | Y | Y |
| CV3 | Express the model completely including all its surrounding elements | Y | Y | Y |
| | *Inter-Model View (IMV)* | | | |
| IMV1 | Dependencies between the models and model elements | Y | Y | Y |
| IMV2 | Indirect dependencies between the models and model elements | Y | Y | Y |
| IMV3 | Integration of the two or more models | N | N | Y |
| | *City View (CiV)* | | | |
| CiV1 | Extendibility of the models | N? | Y | Y |
| CiV2 | Traceability of a model or model element? | N? | Y | Y |
| | *Metric View (MeV)* | | | |
| MeV1 | Metrics to estimate the impact analysis of the models during evolution | NA | Y | NA |
| MeV2 | Visualization techniques for the evolution models | NA | Y | N? |
| MeV3 | Metric values to know the evolution of the models | NA | Y | NA |
| MeV4 | Metrics for quality and complexity of the model | NA | Y? | NA |
| | *Transformation View (TV)* | | | |
| TV1 | Kind of transformation | N? | NA | Y |
| TV2 | Knowledge about the transformation of the models | N? | NA | Y |
| TV3 | Mapping of the models | N? | N | Y? |
| TV4 | Transformation Rules | NA | NA | NA |
| TV5 | Transformation Language | NA | NA | NA |
| | *Evolution View (EV)* | | | |
| EV1 | Trends and causes for evolution of models | Y? | Y | NA |
| EV2 | Multiple dimensions of evolution | N | N? | N? |
| | *Evaluation View (EaV)* | | | |
| EaV1 | Evolution trends and techniques | N | Y | NA |
| EaV2 | Causes for the evolution of models | Y? | Y | NA |
| EaV3 | Stakeholders feedback | Y | N? | N? |

**Dr. Anand Rao Akepogu.** recieved B.Sc (M.P.C) degree from Sri VENKATESWARA University, Andhra Pradesh, India. He received B.Tech degree in Computer Science & Engineering from University of Hyderabad, Andhra Pradesh, India and M.Tech degree in A.I & Robotics from University of Hyderabad, Andhra Pradesh, India. He received PhD degree from Indian Institute of Technology, Madras, India. He is currently working as a Professor & HOD of Computer Science & Engineering Department and also as a Vice-Principal of JNTU College of Engineering, Anantapur, Jawaharlal Nehru technological University, Andhra Pradesh, India. Dr. Rao published more than twenty research papers in international journals and conferences. His main research interest includes software engineering and data mining.

**Madhavi karanam.** recieved B.E degree in Computer Science and Engineering from Kuvempu University, Karanataka, India and M.Tech degree in Software Engineering from Jawaharlal Nehru Technological University, Andhra Pradesh, India. Currrently she is pursuing PhD from Jawaharlal Nehru technological University, Andhra Pradesh, India. Madhavi published six research papers in national and international conferences. Her main research interest includes software visualization, model driven development and software engineering. She is a graduate member of the IEEE Computer Society.